\documentclass[%
 reprint,
 amsmath,amssymb,
 aps,prc
]{revtex4-2}
\usepackage[colorlinks, citecolor=red]{hyperref}
\usepackage{graphicx}
\usepackage{dcolumn}
\usepackage{bm}
\usepackage{nameref}
\usepackage{natbib}
\usepackage[T1]{fontenc}
\usepackage{booktabs, array, mathptmx, float, tabularx, booktabs, lipsum, amsmath,multirow}
\usepackage{siunitx, xcolor}
\usepackage[version=4]{mhchem}

\begin{document}
\preprint{APS/123-QED}
\title{\texorpdfstring{$\alpha$}{}-decay systematics for superheavy nucleus: The effect of deformation of daughter nucleus}

\author{ Jinyu Hu$^{1}$ and  Chen Wu$^{1}$ } \affiliation{
\small 1. Xingzhi College, Zhejiang Normal University, Jinhua, 321004, Zhejiang, China}
\begin{abstract}
Recently, V. Yu. Denisov \cite{denisov2024empirical} introduced quadrupole deformation into the empirical formula for calculating $\alpha$-decay half-lives, leading to a significant improvement in accuracy for even-even nuclei. In this work, we extend this approach by incorporating hexadecapole and hexacontatetrapole deformations into three empirical models: the formula proposed by Deng et al. (DUR)  \cite{deng2020improved}, the formula modified by AKrawy and Poenaru to include nuclear isospin (AKRA) \cite{akrawy2017alpha}, and the improved New Geiger-Nuttall law (NGN) by Y. Ren and Z. Ren \cite{ren2012new}. Using these deformation-enhanced versions-denoted as DUR+D, AKRA+D, and NGN+D-along with their original forms, we calculated the $\alpha$-decay half-lives of 400 isotopes. The results show that AKRA+D achieves the best agreement with experimental data. As an application, we employed DUR+D, AKRA+D, the extended formula by Xu et al. for odd-A nuclei (Improved+UL) \cite{xu2022unified} and odd-odd nuclei (Improved+EF) \cite{luo2023improved}-which account for centrifugal potential, shell effects, and the blocking effect of unpaired nucleons-as well as V. Yu. Denisov's deformation-based empirical formula (ND), to predict $\alpha$-decay properties of 71 even-even nuclei with $\mathit{Z}$ = 118, 120, 122, and 124. Predictions from all five models are in strong agreement, confirming the reliability of our approach and providing valuable guidance for future experiments aimed at synthesizing new elements.
\end{abstract}

\maketitle

\section{\label{sec:level1}INTRODUCTION}
Since Ernest Rutherford first described $\alpha$-decay in 1908 as the emission of a $^{4}$He nucleus from a parent nucleus, it has remained a major research topic in nuclear physics. As the dominant decay mode for superheavy nuclei, $\alpha$-decay provides rich information on nuclear structure-including nuclear spins, shell effects, ground-state energies, and half-lives. Moreover, with progress in experimental detection techniques, $\alpha$-decay chains serve as an important tool for identifying new elements and isotopes. In summary, $\alpha$-decay continues to be a central and active research area in nuclear physics   \cite{qu2014comparative,zdeb2013half,sun2017systematic,santhosh2018alpha,hosseini2019alpha,oganessian2010synthesis,oganessian2007heaviest,hosseini2017theoretical,akrawy2019alpha,hosseini2019alpha,javadimanesh2013investigation,hosseini2019alpha,hodgson2003cluster}.
\par{}
In 1911, Geiger and Nuttall \cite{geiger1911lvii} first introduced an empirical formula to describe alpha decay. This formula suggests that the logarithm of the half-life of alpha decay is linearly correlated with the negative square root of the alpha decay energy. This empirical relationship is known as the Geiger-Nuttall law. It was written as:
\begin{equation}\label{equa1}
log_{10} T_{1/2}(s) = a+bQ_{\alpha}^{-1/2}.
\end{equation}
\par{}
Subsequently, Gamow and, independently, Condon and Gurney interpreted $\alpha$-decay in terms of quantum tunneling, in agreement with the Geiger-Nuttall law. This development motivated a broad range of models for evaluating $\alpha$-decay half-lives, including the Gamow-like model \cite{zdeb2013half}, the modified generalized liquid-drop model \cite{guo2015nuclear,zhang2006alpha}, the Coulomb and proximity potential model \cite{zanganah2020calculation,yahya2020alpha}, the two-potential approach (TPA) \cite{gurvitz1987decay}, and double-folding potential methods \cite{moghaddari2020influence}, among others \cite{jian2009alpha,yan2009branching}. Despite substantial progress, a unified framework capable of describing $\alpha$-decay and providing quantitatively reliable half-life predictions over the full nuclear chart is still lacking. Both the Geiger-Nuttall law and modern theoretical approaches indicate that $\alpha$-decay half-lives are strongly influenced by multiple nuclear-structure and barrier characteristics, which contributes to the persistent challenge of achieving high predictive accuracy.
\par{}
A variety of improved empirical formulas for describing $\alpha$-decay have been developed on the basis of the Geiger-Nuttall law. Early efforts focused on incorporating the mass number $\mathit{A}$, proton number $\mathit{Z}$, and different functional forms of the $\alpha$-decay energy ($\mathit{Q}_{\alpha}$), leading to widely used relations such as the Royer formula \cite{royer2000alpha}, the Viola-Seaborg-Sobiczewski formula \cite{viola1966nuclear}, and the universal decay law (UDL) \cite{qi2009universal}. To provide a more accurate description of $\alpha$-decay, later empirical formulations explicitly included the spin and parity of the ground states of both the parent and daughter nuclei, which notably improved the predicted half-lives of odd-$\mathit{A}$ and odd-odd systems \cite{ren2012new,koura2012phenomenological}. Numerous spin-parity-dependent empirical models have since been proposed \cite{saxena2021new,soylu2021extended,akrawy2022alpha}. More recently, additional refinements have been achieved by incorporating proton-neutron antisymmetry
\cite{akrawy2017alpha,akrawy2018new,akrawy2019alpha,akrawy2022generalization}. In particular, Akrawy and collaborators \cite{akrawy2018new} systematically demonstrated that the inclusion of proton-neutron antisymmetry significantly enhances the predictive accuracy of empirical $\alpha$-decay half-life formulas. Overall, the incorporation of spin-parity effects and proton-neutron antisymmetry has substantially improved the performance of empirical models for $\alpha$-decay half-life calculations.
\par{}
Despite decades of development that have substantially improved the performance of empirical formulas for describing $\alpha$-decay, these approaches still exhibit notable limitations when compared with semi-classical WKB-based theoretical frameworks. For instance, although many theoretical models explicitly account for deformation effects of the daughter nucleus during $\alpha$ emission \cite{yu2009model,ismail2017alpha,deng2017improved,denisov2010erratum}, most empirical formulas neglect such contributions. A significant advancement was achieved in 2024 by V. Yu. Denisov \cite{denisov2024empirical}, who incorporated quadrupole deformation into an empirical $\alpha$-decay formula and systematically compared deformation parameters obtained from three mass models (FRDM, HFB, and WS4 \cite{wang2014surface}) with those from spherical approximations. This modification reduced the root-mean-square deviation of the calculated decimal logarithms of $\alpha$-decay half-lives by approximately 23$\%$. Motivated by Denisov's results-which demonstrate that an accurate treatment of nuclear quadrupole deformation significantly enhances half-life predictions-we extend this approach by including hexadecapole and hexacontatetrapole deformations in the empirical framework. Incorporating these higher-order shape degrees of freedom enables a more realistic description of nuclear deformation and further reduces the discrepancy between calculated and experimental $\alpha$-decay half-lives.
\par{}
Research on superheavy elements has recently attracted considerable attention
\cite{anushree2025entrance,madhu2024cr,manjunatha2018investigations,manjunatha2024effect,sowmya2024optimal}. For example, Manjunatha et al. \cite{manjunatha2018investigations} showed that $\alpha$-decay is the dominant decay mode for the isotopic chain of superheavy nuclei with $\mathit{Z} = 122$. In addition, both quadrupole and hexadecapole deformations are known to play important roles in heavy-ion fusion processes \cite{manjunatha2024effect,sowmya2024optimal}. In this work, building upon the deformation-dependent empirical formulation proposed by V. Yu. Denisov, we extend his approach to several additional empirical models. Specifically, we employ three empirical formulas. First, Yuejiao Ren and Zhongzhou Ren incorporated quantum numbers and the centrifugal potential into the Geiger-Nuttall law, yielding an improved expression known as the NGN formula. Second, D. T. Akrawy and D. N. Poenaru introduced isospin dependence into the Royer formula, resulting in the AKRA relation. Third, Jun-Gang Deng, Hong-Fei Zhang, and G. Royer incorporated the centrifugal potential into the Royer expression, producing the refined DUR model.We apply these three original formulas, together with their deformation-extended versions (DUR+D, AKRA+D, and NGN+D), to analyze the $\alpha$-decay half-lives of 400 selected nuclei. These nuclei are further classified into four categories even-even, even-odd, odd-even, and odd-odd for detailed examination. Importantly, we incorporate both hexadecapole and hexacontatetrapole deformations of the daughter nucleus into the empirical models to investigate how higher-order shape effects improve the predictive accuracy of $\alpha$-decay half-lives. For comparative purposes, we further employ the DUR+D and AKRA+D models, the extended formulas proposed by Xu et al. for odd-$A$ (Improved+UL) and odd-odd (Improved+EF) nuclei-which include centrifugal potential, shell corrections, and blocking effects of unpaired nucleons-as well as Denisov's deformation-based empirical expression (N+D), to calculate the $\alpha$-decay half-lives of 71 even-even nuclei with $118 \leq \mathit{Z} \leq 124$, and subsequently compare their predictive performance.
\par{}
This work is organized as follows. In Sec. \ref{sec:level3}, we apply these three original models and their improved model (the DUR+D model, the AKRA+D model, and the NGN+D model) to evaluate the alpha-decay half-lives for each isotope and compare the results with experimental values. The application of the DUR+D model, the AKRA+D model has been extended to 71 even-even nuclei with  $\mathit{Z}$ = 118 to $\mathit{Z}$ = 124; the result was  shown in figures for each set of isotopes. In Sec. \ref{sec:level4}, we present our conclusions.
\begin{figure*}[ht]
    \centering
    \includegraphics[width=\textwidth]{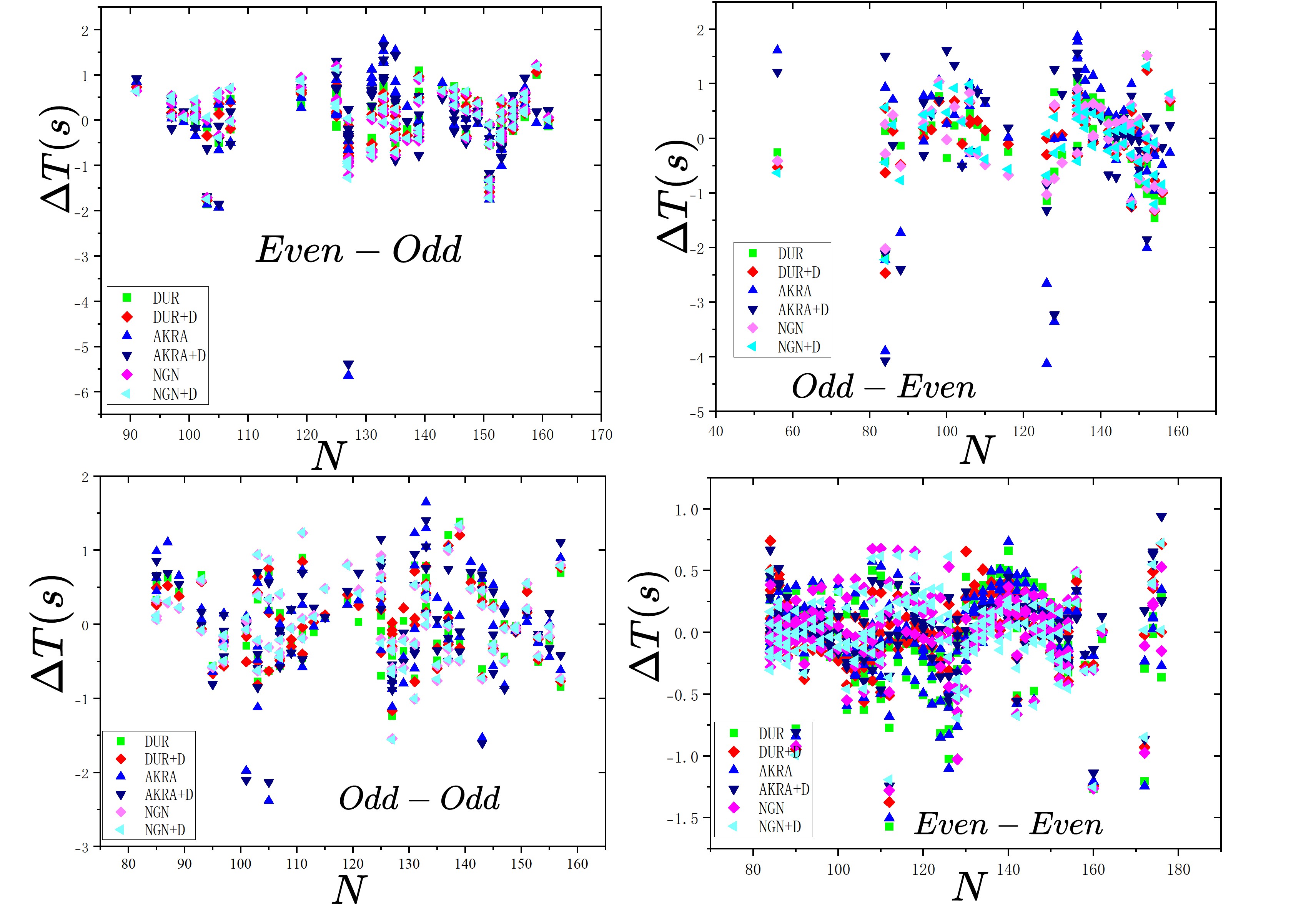}
    \caption{Calculations of $\alpha$ decay half-lives for even-even nuclei, even-odd nuclei, odd-even nuclei and odd-odd nuclei. The experimental $\alpha$ decay half-lives are take from the latest evaluated nuclear properties table NUBASE2020 \cite{kondev2021nubase2020}. The logarithmic differences between calculated and experimental $\alpha$-decay half-lives for the $\alpha$-decay for even-even, even-odd, odd-even, odd-odd nuclei. The calculations are done for the values of the quadrupole, the hexadecapole and the hexacontatetrapole deformation parameters taken from Ref. \cite{wang2014surface}. The values of $\alpha$-decay half-lives are given in seconds.}
    \label{imag1}
\end{figure*}

\section{\label{sec:level2}FORMALISM OF \texorpdfstring{$\alpha$}{}-DECAY HALF-LIVES }
\begin{table*}[ht]
    \renewcommand{\arraystretch}{1}
    \setlength{\tabcolsep}{0.1cm}
    \centering
    \caption{The coefficient of the DUR model (DUR).}
    \begin{ruledtabular}
    \scalebox{1}{
    \begin{tabular}{ccccc}
        Set& a & b&c&d \\
        \colrule
        Even-even &-25.62 & -1.16 &1.60&-\\
        Even-odd & -27.1458 & -1.15 &1.64 &0.045\\
        Odd-even & -27.6654& -1.09 & 1.61&0.054\\
        Odd-odd &-24.9491 & -1.22& 1.62&0.053 \\
    \end{tabular}
    }
    \label{tab1}
    \end{ruledtabular}
\end{table*}
\begin{table*}[ht]
    \renewcommand{\arraystretch}{1}
    \setlength{\tabcolsep}{0.1cm}
    \centering
    \caption{The coefficient of the DUR+D model (DUR+D).}
    \begin{ruledtabular}
    \scalebox{1}{
    \begin{tabular}{cccccc}
        Set& a & b&c&d&e \\
        \colrule
        Even-even & -27.12 &-1.13&1.64&-& -0.035 \\
        Even-odd & -28.70 & -1.12&1.68 &0.044&-0.026\\
        Odd-even & -28.53& -1.07&1.63&0.051&-0.022\\
        Odd-odd & -26.11& -1.16&1.65&0.048&-0.055 \\
    \end{tabular}
    }
    \label{tab2}
    \end{ruledtabular}
\end{table*}
\begin{table*}[ht]
	\renewcommand{\arraystretch}{1}
	\setlength{\tabcolsep}{0.1cm}
	\centering
	\caption{The coefficient of the AKRA model (AKRA).}
	\begin{ruledtabular}
		\scalebox{1}{
			\begin{tabular}{cccccc}
				Set& a & b&c&d&e \\
				\colrule
				Even-even & -26.99 & -1.13&1.61&5.88&-27.02 \\
				Even-odd & -16.57&-1.34&1.44&-6.98&85.06 \\
				Odd-even & -21.62 &-1.24&1.53&8.87&-1.36\\
				Odd-odd & -14.52& -1.33&1.39&-5.45&64.32 \\
			\end{tabular}
		}
		\label{tab3}
	\end{ruledtabular}
\end{table*}
\begin{table*}[ht]
    \renewcommand{\arraystretch}{1}
    \setlength{\tabcolsep}{0.1cm}
    \centering
    \caption{The coefficient of the AKRA+D model (AKRA+D).}
    \begin{ruledtabular}
    \setlength{\tabcolsep}{1pt}
    \begin{tabular}{ccccccc}
        Set& a & b&c&d&e&f \\
        \colrule
        Even-even & -29.51 & -1.06&1.68&-1.21&-13.74&-0.042 \\
        Even-odd & -20.10 & -1.26&1.54&-12.94 &107.30&-0.068\\
        Odd-even & -24.63 & -1.13&1.60&-2.95 &24.53&-0.047\\
        Odd-odd & -18.94 & -1.14&1.52&-22.11 & 104.91&-0.13\\
    \end{tabular}
    \label{tab4}
    \end{ruledtabular}
\end{table*}
\begin{table*}[ht]
    \renewcommand{\arraystretch}{1}
    \setlength{\tabcolsep}{0.05cm}
    \centering
    \caption{The coefficient of the New Geiger-Nuttall law (NGN).}
    \begin{ruledtabular}
    \scalebox{1}{
    \begin{tabular}{ccccc}
        Set& a & b&c&d\\
        \colrule
        Even-even & 0.013&-0.043&-18.13& -\\
        Even-odd & 0.014 & -0.041&-20.25& 0.047\\
        Odd-even & 0.013 &-0.038&-21.45 & 0.060\\
        Odd-odd &  0.013 & -0.045&-16.04& 0.0051\\
    \end{tabular}
    }
    \label{tab5}
    \end{ruledtabular}
\end{table*}

\begin{table*}[ht]
    \renewcommand{\arraystretch}{1}
    \setlength{\tabcolsep}{0.1cm}
    \centering
    \caption{The coefficient of the New Geiger-Nuttall+D law (NGN+D).}
    \begin{ruledtabular}
    \scalebox{1}{
    \begin{tabular}{cccccc}
        Set& a & b&c&d&e \\
        \colrule
        Even-even & 0.014 & -0.043& -18.89 & - & -0.016\\
        Even-odd & 0.0136 & -0.041& -19.64& 0.047& 0.00902\\
        Odd-even & 0.0134& -0.038& -21.26 & 0.060 & 0.00416\\
        Odd-odd & 0.014& -0.044& -17.16& 0.047& -0.042\\
    \end{tabular}
    }
    \label{tab6}
    \end{ruledtabular}
\end{table*}
\par{}
Recently, V. Yu. Denisov \cite{denisov2024empirical} proposed a new empirical formula by introducing the deformation of the daughter nucleus. The new empirical relation is given as
\begin{equation}\label{equa2}
\begin{aligned}
log_{10}T_{1/2}(s) = & a\frac{Z}{Q^{1/2}} - b(\frac{AQ^{1/2}}{Z})^{1/6}-cA^{1/6}Z^{1/2} \\&
                   +dA^{1/6}\frac{\sqrt{l(l+1)}}{Q} - e(k\mathit{\mathit{\mathit{\beta}}})^{1/2}\frac{Z}{Q^{1/2}}, 
\end{aligned}
\end{equation}
where the half-life is given in seconds, the $\alpha$ decay energy ($\mathit{Q}$) in MeV, A is the mass number of parent nucleus, Z is the proton number of parent nucleus, and $\mathit{\beta}$ is the quadrupole deformation parameter of the deformed daughter nucleus.
\par{}
V.Yu.Denisov's study introduces a fifth term in the $\alpha$-decay half-life calculation, which accounts for the influence of the daughter nucleus's deformation on the Coulomb interaction between the emitted $\alpha$-particle and the daughter nucleus. This effect is crucial because the half-life is significantly influenced by the daughter nucleus's deformation. Denisov defines the effect of daughter nucleus deformation on the $\alpha$-decay half-life based on the minimum value of the Coulomb interaction, $V^{min}_{C}$. This minimum value is expressed as:
\begin{equation}\label{equ1}
\begin{aligned}
V^{min}_{C} = 2(Z-2)e^{2}_{p}/(R_{L}+R_{\alpha}), 
\end{aligned}
\end{equation}
where $\mathit{Z}$ is the proton number of the parent nucleus, $e_{p}$ is the proton charge, $R_{\alpha}$ is the radius of the $\alpha$-particle, and $R_{L}$ is the largest radius of the deformed daughter nucleus.
\par{}
The most critical parameter governing the Coulomb interaction is the maximum value of the daughter nucleus's quadrupole deformation parameter, $\mathit{\beta}$. The surface radius of the deformed daughter nucleus, $R(\theta)$, is typically modeled by the expression $R(\theta) = R_{0}[1+\mathit{\beta} Y_{20}(\theta)]$, where $R_{0}$ is the radius of the corresponding spherical daughter nucleus. The value of $R_{L}$ depends on the sign of the deformation parameter $\mathit{\beta}$:
For prolate deformation ($\mathit{\beta} > 0$): The largest radius occurs at $\theta= 0$ (the poles), and is given by $R_{L} = R_{0}(1+\sqrt{5/\pi}\mathit{\beta}/2)$.For oblate deformation ($\mathit{\beta} < 0$): The largest radius occurs at $\theta= \pi/2$ (the equator), and is given by $R_{L} = R_{0}(1+\sqrt{5/\pi}\mathit{\beta}/4)$. The reduction in the $\alpha$-decay half-life caused by the daughter nucleus's deformation is quantified by subtracting the Coulomb interaction of the spherical nucleus from the minimum Coulomb interaction ($V^{min}_{C}$) induced by the deformation. In summary, V. Yu. Denisov's analysis demonstrates that the reduction in the $\alpha$-decay half-life attributed to daughter nucleus deformation is proportional to the difference between the largest deformed radius and the spherical radius ($\Delta \propto R_{L}-R_{0}$).
\par{}
V. Yu. Denisov's research established that the alteration in the $\alpha$-decay half-life due to daughter nucleus deformation can be characterized by changes in the Coulomb interaction between the $\alpha$-particle and the daughter nucleus. However, Denisov's original formulation considered only quadrupole deformation ($\mathit{\beta}_2$).In this work, we extend this framework by incorporating hexadecapole ($\mathit{\beta}_4$) and hexacontatetrapole ($\mathit{\beta}_6$) deformations of the daughter nucleus. The minimum value of the Coulomb interaction, $V^{min}_{C}$, remains defined based on the maximum distance:$V^{min}_{C} = 2(Z-2)e^{2}_{p}/(R_{L}+R_{\alpha})$. With the inclusion of higher-order terms, the radius of the deformed daughter nucleus, $R$, is given by: $$R(\theta) = R_{0}(1+\mathit{\beta}_{2}Y_{20}(\theta)+\mathit{\beta}_{4}Y_{40}(\theta)+\mathit{\beta}_{6}Y_{60}(\theta)), $$  where $R_{0}$ is the radius of the spherical daughter nucleus and $Y_{l0}(\theta)$ are spherical harmonics.To comprehensively describe the characteristics of the deformed daughter nucleus relevant to the $\alpha$-particle's tunneling probability, it is essential to consider the maximum value of its radius, $R_{L}$. This maximum radius is determined by finding the largest positive contribution from the deformation terms:
\begin{equation}\label{equa4}
\begin{aligned}
R_{L} = R_{0}(1+\left | Max(\mathit{\beta}_{2}Y_{20}(\theta)+\mathit{\beta}_{4}Y_{40}(\theta)+\mathit{\beta}_{6}Y_{60}(\theta)) \right | ).
\end{aligned}
\end{equation}
Since this deformation primarily influences the Coulomb interaction, which is a key factor in the $\alpha$-decay barrier height, we deduce a new deformation term for empirical half-life formulas. Based on an analysis informed by the Royer formula (which relates half-life to the Coulomb barrier height and $\alpha$-decay energy, $Q$), the deformation term incorporated into the empirical half-life calculation is expressed as:
\begin{equation}\label{equa5}
\begin{aligned}
e(\left | Max(\mathit{\beta}_{2}Y_{20}(\theta)+\mathit{\beta}_{4}Y_{40}(\theta)+\mathit{\beta}_{6}Y_{60}(\theta)) \right |)^{1/2}\frac{Z}{Q^{1/2}}.
\end{aligned}
\end{equation}
\par{}
We incorporated this new deformation term (Eq. \ref{equa5}) into three established empirical formulas: the DUR model, the AKRA model, and the NGN model. This modification yields the improved versions: DUR+D, AKRA+D, and NGN+D.Subsequently, we used both the original and the modified versions (a total of six empirical formulas) to calculate the $\alpha$-decay half-lives for a selected dataset of 400 nuclei. The necessary parameter fitting and curve fitting analysis were performed using the dedicated modules within the Python 3.11 programming language environment.
\par{}
Prior to presenting our analysis, it is essential to briefly review the six phenomenological models that underpin this study.
\par{}

 \subsection{The DUR model(DUR)}\label{A1}
\par{}
  In 2020, Deng et al. \cite{deng2020improved} presented an unitary Royer formula(DUR) for $\alpha$ decay half-lives. It can be expressed as
\begin{equation}\label{equa6}
log_{10}T_{1/2} = a +bA^{1/6}\sqrt{Z}+cZ/\sqrt{Q_{\alpha}}+dl(l+1)+h,
\end{equation}
where the half-life is given in seconds, and decay energy(Q) in MeV, $A$ and $Z$ are the mass and proton numbers of parent nucleus,respectively. The optimal parameters for $a$, $b$, $c$,and $d$ were obtained through fitting to experimental data and are listed in Table \ref{tab1}. The h is given by
\begin{equation}\label{equa7}
h=\left\{\begin{matrix}
  0,& for\:even\:-\:even\:nucleus,\\
  0.2812& for\:odd\:Z\:-\:even\:N\:nucleus,\\
  0.3625& for\:even\:Z\:-\:odd\:N\:nucleus,\\
  0.7486&for\:odd\:-\:odd\:nucleus.
\end{matrix}\right.
\end{equation}

 \subsection{The DUR+D model(DUR+D)}\label{B2}
\par{}
  In the present work, we modify the DUR model by adding deformation terms,and the DUR+D model takes the form
\begin{equation}\label{equa8}
\begin{aligned}
log_{10}T_{1/2}(s) = & a +bA^{1/6}\sqrt{Z}+cZ/\sqrt{Q_{\alpha}}+dl(l+1)+h   \\&
                   +e(\left | Max(\mathit{\beta}_{2}Y_{20}(\theta)+\mathit{\beta}_{4}Y_{40}(\theta)+\mathit{\beta}_{6}Y_{60}(\theta)) \right |)^{1/2}\\&
                   \times\frac{Z}{Q^{1/2}},
\end{aligned}
\end{equation}
where $\mathit{\beta}_{2}$, $\mathit{\beta}_{4}$, and $\mathit{\beta}_{6}$ are the quadrupole, hexadecapole and hexacontatetrapole deformation of the deformed daughter nucleus, while the optimal parameters $a$, $b$, $c$, $d$, $e$ are the obtained fitting to experimental data and are listed in Table \ref{tab3}.
 \subsection{The AKRA model(AKRA)}\label{C3}
\par{}
   Akrawy and Poenaru \cite{akrawy2017alpha} reported a new relation a new relationship for the calculations of alpha decay half-lives by introducing iso-spin asymmetry, $I$, which is based on the Royer relationship. The new semiempirical relationship is given as,
\begin{equation}\label{equa9}
log_{10}T_{1/2}(s) = a+b A^{1/6}\sqrt{Z}+ c \frac{Z}{\sqrt{Q}}+dI+eI^{2},
\end{equation}
where $I$ is the asymmetry term, $I=\frac{N-Z}{A}$. The optimal parameters for $a$, $b$, $c$, $d$, $e$ were obtained through fitting to experimental data and are listed in Table \ref{tab3}.
 \subsection{the AKRA+D model (AKRA+D)}\label{D4}
\par{}
   The modification of the AKRA model was also done by adding the deformation terms; The AKRA+D model is given as,
\begin{equation}\label{equa10}
\begin{aligned}
log_{10}T_{1/2}(s) = & a+b A^{1/6}\sqrt{Z}+ c \frac{Z}{\sqrt{Q}}+dI+eI^{2}  \\&
                   +f(\left | Max(\mathit{\beta}_{2}Y_{20}(\theta)+\mathit{\beta}_{4}Y_{40}(\theta)+\mathit{\beta}_{6}Y_{60}(\theta)) \right |)^{1/2}\\&
                   \times\frac{Z}{Q^{1/2}},
\end{aligned}
\end{equation}
where $\mathit{\beta}_{2}$, $\mathit{\beta}_{4}$, and $\mathit{\beta}_{6}$ are the quadrupole, hexadecapole and hexacontatetrapole deformation of the deformed daughter nucleus, while the optimal parameters $a$, $b$, $c$, $d$, $e$, $f$ re the obtained fitting to experimental data and are listed in Table \ref{tab4}.

 \subsection{The New Geiger-Nuttall law (NGN)}\label{E5}
\par{}
   The New Geiger-Nuttall law formula (NGN) \cite{ren2012new} for \texorpdfstring{$\alpha$}{}-decay half-lives is written as
\begin{equation}\label{equa11}
\begin{aligned}
log_{10}T_{1/2}(s) = & a\frac{\mu Z_{c}Z_{d}}{\sqrt{Q}}+b\sqrt{\mu}\sqrt{Z_{c}Z_{d}}   \\&
                   +c+S+dl(l+1).
\end{aligned}
\end{equation}
Here $Q$ is the decay energy in MeV units, $Z_{c}$ and $Z_{d}$ are the charge number of the $\alpha$ particle and the daughter nucleus, $S$ = 0 for $N\ge127$ and $S$ = 1 for $N\le 126$, and $\mu$ is reduced mass the respectively, and the optimal parameters for $a$, $b$ and $c$, $d$ are the obtained fitting to experimental data and are listed in Table \ref{tab5}.

 \subsection{The New Geiger-Nuttall+D law (NGN+D)}\label{F6}
\par{}
    Similarly, as we did with the DUR model, the New Geiger-Nuttall law has been modified adding the deformation terms; the New Geiger-Nuttall+D (NGN+D) will be
\begin{equation}\label{equa12}
\begin{aligned}
log_{10}T_{1/2}(s) =  & a\frac{\mu Z_{c}Z_{d}}{\sqrt{Q}}+b\sqrt{\mu}\sqrt{Z_{c}Z_{d}}   \\&
                   +c+S+dl(l+1) \\&
                   +e(\left | Max(\mathit{\beta}_{2}Y_{20}(\theta)+\mathit{\beta}_{4}Y_{40}(\theta)+\mathit{\beta}_{6}Y_{60}(\theta)) \right |)^{1/2}\\&
                   \times\frac{Z}{Q^{1/2}},
\end{aligned}
\end{equation}
where $\mathit{\beta}_{2}$, $\mathit{\beta}_{4}$, and $\mathit{\beta}_{6}$ are the quadrupole, hexadecapole and hexacontatetrapole deformation of the deformed daughter nucleus, while the optimal parameters for $a$, $b$, $c$, $d$, $e$  are the obtained fitting to experimental data and are listed in Table \ref{tab6}.

\section{\label{sec:level3}RESULTS AND DISCUSSION}
In this study, we partitioned 400 nuclei into four groups (even-even, even-odd, odd-even, and odd-odd) and employed six empirical formulas-the DUR model, the AKRA model, the New Geiger-Nuttall (NGN) formula, and their deformation-extended versions (DUR+D, AKRA+D, and NGN+D)-to calculate their $\alpha$-decay half-lives. V. Yu. Denisov demonstrated that the WS4 deformation parameters $\mathit{\beta}_{2}$, $\mathit{\beta}_{4}$, and $\mathit{\beta}_{6}$ yield the smallest deviations between calculated and experimental $\alpha$-decay half-lives; thus, these deformation parameters are adopted in the present work. To enable a comprehensive comparison of the performance of the six models, we evaluate their predictive accuracy using the root-mean-square (RMS) deviation between the calculated and measured half-lives. The (RMS) is defined as
\begin{equation}\label{equa13}
\sigma = \left \{ \frac{1}{n} \sum_{i=1}^{n}\left [ log_{10}(T_{\frac{1}{2},i}^{calc.})-log_{10}(T_{\frac{1}{2},i}^{expt.}) \right ]^{2} \right \} ^{1/2},
\end{equation}
 where $log_{10}(T_{\frac{1}{2},i}^{calc.})$ and $log_{10}(T_{\frac{1}{2},i}^{expt.})$ are the calculated and experimental \texorpdfstring{$\alpha$}{}-decay half-lives the nucleus, and $n$ is the number of nucleus involved for each group.
\par{}
The calculated RMS for each model is listed in Table \ref{tab7}. The results shows the superiority the AKRA+D model over the other five models for all the studied sets of nucleus.
\par{}
Another approach to compare the performance of different models is to calculate the difference between the calculated results and experimental data for each model. This can be expressed as:
\begin{equation}\label{equa14}
\Delta T = log_{10}(T_{\frac{1}{2},i}^{calc.})-log_{10}(T_{\frac{1}{2},i}^{expt.}).
\end{equation}
\par{}
As shown in Table \ref{tab8}, the AKRA+D model reduces the deviation between calculated and experimental $\alpha$-decay half-lives by 22$\%$ for even-even nuclei, 5$\%$ for odd-odd nuclei, 4.7$\%$ for odd-even nuclei, and 14.7$\%$ for another odd-odd subset. These results indicate that the AKRA+D model outperforms the other five models considered, as well as the recently proposed deformation-dependent empirical formula (ND) of V. Yu. Denisov. Figure \ref{imag1} further reveals pronounced discontinuities between the calculated and experimental values for all models at the neutron shell closures $N=126$ and $N=152$. The first discontinuity corresponds to the well-established magic number $\mathit{N} = 126$, which strongly influences the preformation probability of the $\alpha$ particle. The second discontinuity provides compelling evidence supporting $\mathit{N} = 152$ as the next neutron magic number. Our findings are consistent with the conclusions reported by Yuejiao Ren and Zhongzhou Ren \cite{ren2012new}.
\par{}
Three principal empirical models are examined in this work. The DUR model refines the Royer formula by incorporating the Coulomb interaction and accounting for the blocking effect of unpaired nucleons. The NGN model extends the Geiger-Nuttall law by introducing a Coulomb potential together with shell-effect quantum numbers. The AKRA model further modifies the Royer formula by adding an explicit isospin dependence. In heavy and superheavy nuclei, both deformation and isospin effects become significantly enhanced, indicating a potential synergistic relationship between them. Moreover, the work of V. Yu. Denisov shows that, for even-even nuclei, a precise treatment of nuclear spin-parity properties can markedly improve deformation-dependent empirical formulations. These findings together suggest that empirical formulas incorporating both nuclear deformation and spin-parity effects are better suited for accurately reproducing $\alpha$-decay half-lives. Among the deformation-extended models considered here, only the AKRA+D formulation simultaneously includes nuclear deformation and spin-parity effects, whereas DUR+D and NGN+D consistent with the observations of Denisov include only the quantum-number contributions associated with unpaired-nucleon blocking and shell effects, respectively. Our results clearly indicate that, for heavy and superheavy nuclei, isospin plays a more critical role than the other factors. We also plot the resulting $\Delta T$ values as a function of mass number for even-even, even-odd, odd-even, and odd-odd nuclei (Fig. \ref{imag1}). The superiority of the AKRA+D model is evident: its deviations are systematically smaller and exhibit smoother behavior across the entire nuclear chart. In particular, the deviation patterns for odd-$A$ systems shown in Fig. \ref{imag1} demonstrate that incorporating both nuclear asymmetry and daughter-nucleus deformation yields predictions in excellent agreement with experimental data.
\par{}
The strong correlation between Coulomb barrier parameters and different orders of nuclear deformation, as reported by Ismail et al. \cite{ismail2009effect}, aligns well with the foundational research by V. Yu. Denisov \cite{denisov2024empirical}. Denisov initially quantified the influence of daughter nucleus deformation on the Coulomb energy of the $\alpha$-daughter system by considering the quadrupole deformation ($\mathit{\beta}_2$). He proposed using the difference between the maximum radius of the deformed nucleus ($R_{L}$) and the radius of the corresponding spherical nucleus ($R_{0}$) as a measure of this influence ($\propto R_{L}-R_{0}$). Building upon this, our work focuses on the three dominant deformation parameters: quadrupole ($\mathit{\beta}_2$), hexadecapole ($\mathit{\beta}_4$), and hexacontatetrapole ($\mathit{\beta}_6$). This selection is justified by the understanding that the influence of higher-order deformations (beyond $\mathit{\beta}_6$) is typically suppressed by the larger $\mathit{\beta}_2$ and $\mathit{\beta}_4$ values. Crucially, studies have confirmed the importance of these higher-order terms: The work of M. Ismail et al. \cite{ismail2009effect} specifically indicates that the hexacontatetrapole deformation ($\mathit{\beta}_6$) directly influences the Coulomb barrier height, causing variations on the order of $1 \text{ MeV}$. Qiong Xiao et al. \cite{xiao2023half} supported the inclusion of deformation factors in semi-classical formulas by demonstrating that incorporating $\mathit{\beta}_2$, $\mathit{\beta}_4$, and $\mathit{\beta}_6$ into the square well radius, $R_{in}(\theta)$, results in a higher tunneling potential peak for deformed nuclei. Furthermore, the research by Narayanaswamy Manjunatha et al. \cite{manjunatha2024effect} on fusion reactions suggests that $\mathit{\beta}_2$ enhances the reaction cross-section between helium and uranium, whereas $\mathit{\beta}_4$ reduces it. This differential effect strongly suggests that higher-order multipole deformations significantly influence the preformation probability of the $\alpha$-particle within the parent nucleus. Given this compelling evidence, we enhance three established empirical $\alpha$-decay models (AKRA, DUR, and NGN) by incorporating a unified deformation term that accounts for $\mathit{\beta}_2$, $\mathit{\beta}_4$, and $\mathit{\beta}_6$. This approach aims to accurately describe the complex interplay between nuclear shape and the $\alpha$-decay probability.
\begin{figure*}[b]
    \centering
    \includegraphics[width=\textwidth]{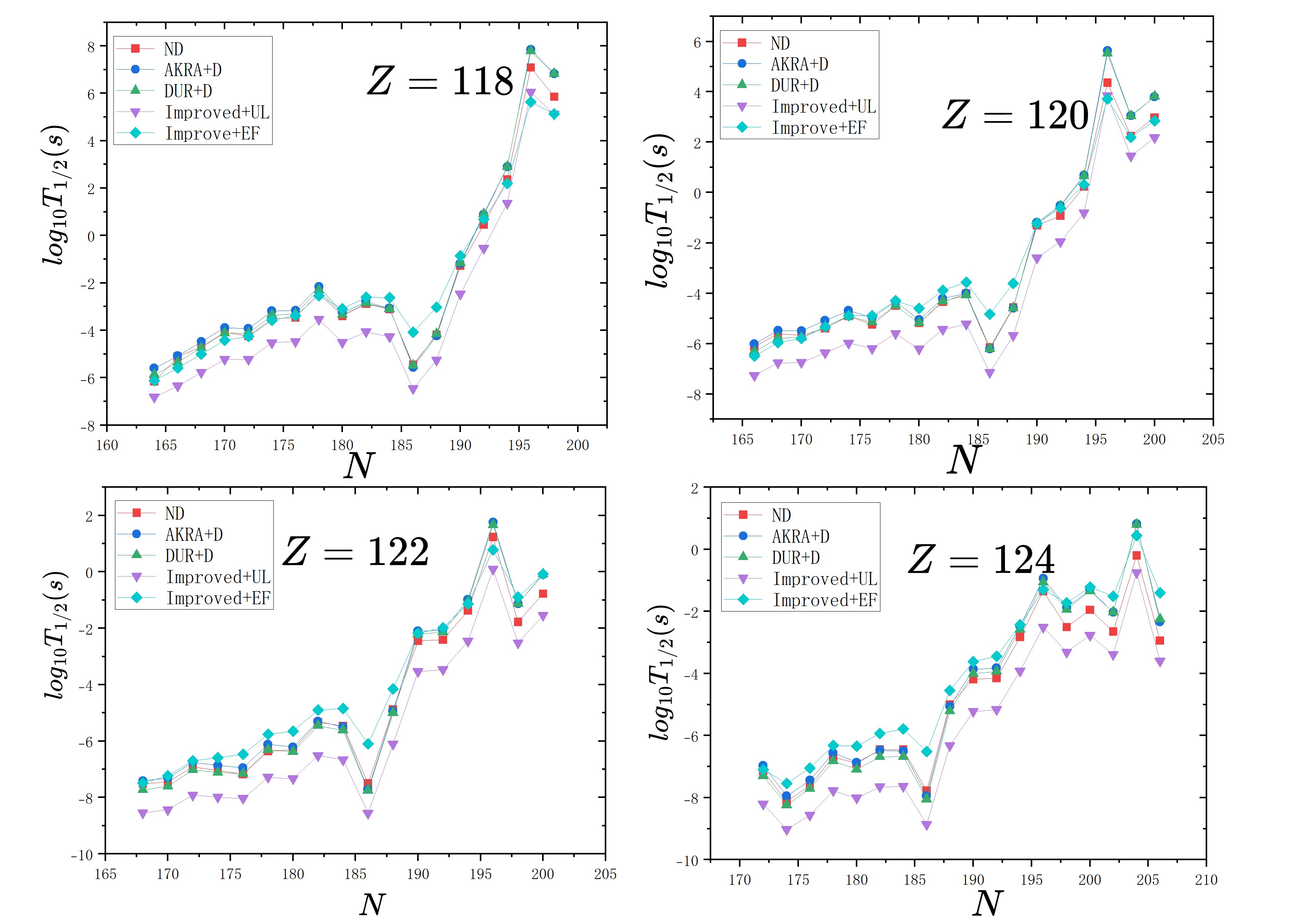}
    \caption{The predicted $\alpha$-decay half-lives in logaruthmic form of even-even nuclei with $Z = 118, 120, 122$, and 124 using Eq. (\ref{equa7}) and Eq. (\ref{equa9}), ND, Improved+UL, and Improved+EF with $Q_{\alpha}$ obtained by WS4 \cite{wang2014surface}. The red square, the blue circle, the green triangle, the purple triangles and cyan diamonds denote the predictions by ND, Eq. (\ref{equa7}), Eq. (\ref{equa9}), Improved+UL, and Improved+EF respectively.}
    \label{imag2}
\end{figure*}
\par{}
For a comprehensive comparative analysis, we employed five empirical models to predict the $\alpha$-decay half-lives of 71 even-even nuclei spanning $\mathit{Z} = 118, 120, 122$, and $124$. The models used were the AKRA+D model, the DUR+D model, the extended formula by Xu et al. for odd-A nuclei (Improved+UL) and odd-odd nuclei (Improved+EF)-which account for centrifugal potential, shell effects, and the blocking effect of unpaired nucleons-as well as V. Yu. Denisov's deformation-based empirical formula (ND). To ensure consistency in the predictions, the essential input parameters-the $\alpha$-decay energy ($\mathit{Q_{\alpha}}$) and the quadrupole ($\mathit{\beta}_2$) deformation parameters of the daughter nuclei-were systematically obtained from the WS4 mass table \cite{wang2014surface}. The complete set of prediction results is summarized in Figure \ref{imag2}, which visually illustrates the computational outcomes of each model. The predictions generated by the AKRA+D and DUR+D models demonstrate substantial agreement with those from the ND model, the Improved+UL model, and the Improved+EF model across the studied isotopic chains. A notable observation is the slight upward deviation exhibited by the predictions of the AKRA+D and DUR+D models compared to the ND model as the neutron number ($\mathit{N}$) increases. This divergence can be directly attributed to the fundamental difference in model formulation: the AKRA+D and DUR+D models explicitly incorporate the contributions from hexadecapole ($\mathit{\beta}_4$) and hexacontatetrapole ($\mathit{\beta}_6$) deformations, which are not accounted for in the ND model. Furthermore, the predicted $\alpha$-decay half-lives for the 71 even-even nuclei show distinct and consistent behavior at the parent neutron numbers $\mathit{N} = 180$ and $\mathit{N} = 186$. This systematic change suggests that the daughter neutron numbers corresponding to $\mathit{N_{daughter}} = 178$ and $\mathit{N_{daughter}} = 184$ may indicate the presence of a neutron magic number and a neutron submagic number, respectively, influencing the nuclear structure and decay dynamics of the superheavy region.
\par{}
In this study, we successfully extended the approach pioneered by V. Yu. Denisov by incorporating hexadecapole ($\mathit{\beta}_4$) and hexacontatetrapole ($\mathit{\beta}_6$) deformations into three classical semi-empirical formulas for $\alpha$-decay half-life calculations. Among the resulting modified models (AKRA+D, DUR+D, NGN+D), the AKRA+D model demonstrated the best overall performance in predicting $\alpha$-decay half-lives. Despite its superior predictive power, the AKRA+D model, in its current form, presents two significant limitations that warrant further investigation: The model successfully accounts for the changes in the Coulomb interaction that arise from the deformation of the daughter nucleus. However, it does not include the corresponding changes in the $\alpha$-decay energy ($Q_{\alpha}$) that are inherently caused by the same nuclear deformation. A comprehensive approach should ideally couple the deformation-induced changes in both the Coulomb barrier and the nuclear mass (which determines $Q_{\alpha}$) for maximum accuracy. While higher-order deformations ($\mathit{\beta}_4$ and $\mathit{\beta}_6$) have been formally included in the model's structure, the available deformation parameters for nuclei remain insufficiently accurate or refined. This uncertainty in the input deformation parameters ultimately limits the ability of the AKRA+D model to fully and precisely describe the complex $\alpha$-decay behavior, particularly in the region of superheavy nuclei.Future work should focus on developing models that internally link the deformation effects on both the Coulomb barrier and the $\alpha$-decay $Q$ value, while also leveraging or generating more precise deformation parameters.
\section{\label{sec:level4}SUMMARY AND CONCLUSION}
\par{}
Building upon the foundational work of V. Yu. Denisov, a new empirical formula incorporating deformation terms (ND) was proposed. We generalized these deformation terms to three classical semi-empirical frameworks: the DUR model, the AKRA model, and the New Geiger-Nuttall law (NGN), leading to the modified models: DUR+D, AKRA+D, and NGN+D. These six empirical formulas (the three original and three modified versions) were systematically employed to investigate the $\alpha$-decay half-lives of a comprehensive set of 400 nuclei, spanning all four parity types (even-even, even-odd, odd-even, and odd-odd). The model performance was quantitatively assessed using the root-mean-square (RMS) deviation between the calculated and experimental $\alpha$-decay half-lives. As illustrated in Figure \ref{imag1}, among the six models tested, the modified AKRA model (AKRA+D) consistently demonstrated the closest agreement with experimental results, indicating its superior predictive capability across the diverse nuclear dataset. The specific reduction in the RMS deviation achieved by the AKRA+D model compared to its original counterpart is detailed in Table \ref{tab8}: The discrepancy between calculated and experimental $\alpha$-decay half-lives was reduced by $22\%$ in the even-even Nuclei. The discrepancy was reduced by $5\%$ in odd-odd Nuclei. The discrepancy was reduced by $4.7\%$ in the odd-even nuclei. The discrepancy was reduced by $14.7\%$ in odd-odd nuclei. These results emphatically confirm that the inclusion of deformation terms, particularly within the AKRA framework, significantly enhances the accuracy of $\alpha$-decay half-life predictions across various nuclear types.
\par{}
To further validate the improvements achieved by incorporating higher-order deformation, we employed a set of five advanced empirical formulas to predict the $\alpha$-decay half-lives of 71 even-even nuclei spanning the superheavy elements $\mathit{Z} = 118, 120, 122$, and $124$.The models used for this predictive study include the deformation-incorporated models developed in this work (the DUR+D model and the AKRA+D model), the new deformation formula by Denisov (ND), and two extended formulas by Xu et al. \cite{xu2022unified,luo2023improved}, which account for centrifugal potential, shell effects, and the blocking effect of unpaired nucleons: Improved+UL (originally for odd-$\mathit{A}$ nuclei) and Improved+EF (originally for odd-odd nuclei). As demonstrated in Figure \ref{imag2}, a strong consistency is observed among the predictions from these five models. Crucially, the predicted $\alpha$-decay half-lives for the 71 even-even nuclei exhibit distinct and consistent behavior at the parent neutron numbers $\mathit{N} = 180$ and $\mathit{N} = 186$. This systematic stability change suggests that the corresponding daughter nucleus neutron numbers, $\mathit{N} = 178$ and $\mathit{N} = 184$, may correspond to a neutron magic number and a neutron submagic number, respectively, providing critical insight into the nuclear structure in the superheavy region. A key finding is the divergence of predictions at larger neutron numbers: for $\mathit{N} > 190$, the half-life predictions from the DUR+D model and the AKRA+D model consistently exceed those of the ND formula.This increasing divergence is directly attributed to the inclusion of hexadecapole ($\mathit{\beta}_4$) and hexacontatetrapole ($\mathit{\beta}_6$) deformation contributions in the DUR+D and AKRA+D models, terms which are not considered in the ND model. This result strongly validates the conclusion that incorporating these higher-order daughter nucleus deformation terms into empirical formulas significantly enhances the accuracy and physical reliability of $\alpha$-decay half-life calculations, particularly for highly deformed even-even nuclei.

\begin{table*}[b]
    \renewcommand{\arraystretch}{1}
    \setlength{\tabcolsep}{0.1cm}
    \centering
    \caption{The RMS deviation of the models DUR, DUR+D, AKRA, AKRA+D,NGN,and NGN+D.}
    \begin{ruledtabular}
    \scalebox{1}{
    \begin{tabular}{ccccc}
        Formula& Even-even n=181& Even-odd n=79&Odd-even n=80&Odd-odd n=60 \\
        \colrule
        DUR & 0.3627 &0.5565&0.4874&0.6595\\
        DUR+D & 0.3040 & 0.5449&0.4766&0.6016\\
        The RMS reduction ($\%$).&16&2&2.2&8.7\\
        AKRA & 0.3555 & 0.9596&0.7011&1.2580\\
        AKRA+D &0.2775 & 0.9114&0.6676&1.077 \\
        The RMS reduction ($\%$).&22&5&4.7&14.4\\
        NGN & 0.3121& 0.6250&0.5213&0.6492\\
        NGN+D & 0.2994& 0.6237&0.5209&0.6165\\
        The RMS reduction ($\%$).&4&0.2&0.07&5\\
    \end{tabular}
    }
    \label{tab7}
    \end{ruledtabular}
\end{table*}
\begin{table*}[b]
    \renewcommand{\arraystretch}{1}
    \setlength{\tabcolsep}{0.1cm}
    \centering
    \caption{$\Delta T$ different between  experimental and theoretical formulas.}
    \begin{ruledtabular}
    \scalebox{1}{
    \begin{tabular}{ccccccccc}
       \multirow{2}{*}{Formula} &\multicolumn{2}{c}{ Even-even} & \multicolumn{2}{c}{ Even-odd} &\multicolumn{2}{c}{ Odd-even} &\multicolumn{2}{c}{ Odd-odd}  \\
              \cmidrule(lr){2-3}\cmidrule(lr){4-5}\cmidrule(lr){6-7}\cmidrule(lr){8-9}
              & Minimum&Maximum & Minimum&Maximum & Minimum&Maximum & Minimum&Maximum \\
        \colrule
        DUR & -1.5708 & 0.6605&-1.8560&1.0952 & -1.2388& 1.3882&-2.1975&1.5149\\
        DUR+D &  -1.2194 & 0.7444&-1.7726&1.0750& -1.1644 &1.2056&-2.4669&1.2479 \\
        AKRA & -1.5031 &  0.7323&-5.6337&1.7573& -2.3818& 1.6460&-4.1263&1.8640 \\
        AKRA+D &-1.1861 & 0.7944&-5.3777& 1.6525 & -2.1315&1.4041&-4.073&1.6130\\
        NGN &  -1.2795 & 0.6773&-1.7121&1.2159& -1.5413& 1.3080&-2.022&1.5198 \\
        NGN+D & -1.2485 &0.6930&-1.7423& 1.1905& -1.5554 & 1.3428&-2.22071&1.3167 \\
    \end{tabular}
    }
    \label{tab8}
    \end{ruledtabular}
\end{table*}
\begin{table*}[b]
    \renewcommand{\arraystretch}{1}
    \setlength{\tabcolsep}{0.1cm}
    \centering
    \caption{The DUR model $log_{10}T_{AKRA+D}$ vs $log_{10}T_{DUR+D}$ vs $log_{10}T_{ND}$. Values for $Q_{\alpha}$ is from Refs. \cite{wang2014surface}.}
    \begin{ruledtabular}
    \scalebox{1}{
    \begin{tabular}{lccccccccccc}
        Z& A& $Q_{\alpha}$&$log_{10}T_{AKRA+D}$&$log_{10}T_{DUR+D}$&$log_{10}T_{ND}$& Z& A& $Q_{\alpha}$&$log_{10}T_{AKRA+D}$&$log_{10}T_{DUR+D}$&$log_{10}T_{ND}$ \\
        \colrule
        118& 282 & 13.492&-5.598&-5.876&-6.160&122& 290& 15.092&-7.414&-7.731&-7.514\\
        &284 & 13.209&-5.084&-5.345&-5.116 && 292& 14.994&-7.309&-7.600&-7.450\\
        & 286 & 12.889&-4.485&-4.729&-4.704 && 294 & 14.643&-6.747&-7.022&-6.922\\
          &288 &  12.587&-3.893&-4.120&-4.079&& 296 &14.670&-6.866&-7.108&-7.038\\
         & 290 &  12.572&-3.931&-4.126 &-4.250&& 298 & 14.678&-6.945&-7.156&-7.179\\
         & 292 &  12.212&-3.182& -3.363 &-3.508&& 300 & 14.197&-6.114&-6.314&-6.353\\
        & 294 & 12.171&-3.156&-3.305 &-3.459&& 302 & 14.212&-6.210&-6.378&-6.307\\
         & 296 & 11.726&-2.157&-2.298&-2.421&& 304 & 13.714&-5.299&-5.457&-5.333\\
          &298& 12.158&-3.264&-3.345&-3.399&& 306 & 13.780&-5.499&-5.621&-5.463\\
         & 300 & 11.932&-2.797&-2.857&-2.876&& 308 & 14.918&-7.715&-7.753&-7.498\\
         & 302 & 12.018&-3.072& -3.092&-3.101 && 310 & 13.435&-4.931&-5.004&-4.874\\
         & 304 &  13.101&-5.557&-5.484&-5.444 && 312 & 12.141&-2.102&-2.209&-2.445\\
         & 306 & 12.459&-4.235&-4.160&-4.192 && 314 & 12.096&-2.061&-2.136&-2.408\\
          &308 &11.184&-1.191&-1.155&-1.282&& 316 & 11.638&-0.982&-1.050&-1.372\\
         & 310 &  10.414&0.886&0.908 &0.461&& 318 & 10.619&1.762&1.662&1.238\\
         & 312 & 9.742&2.897&2.906&2.368&& 320 & 11.637&-1.128&-1.125&-1.778\\
      & 314 &  8.365&7.854&7.780&7.088&& 322 & 11.221&-0.093&-0.082&-0.773\\
        & 316 &  8.601&6.829&6.815&5.854&124& 296 & 15.117&-6.965&-7.304&-7.152\\
     120& 286 & 14.013&-6.012&-6.322&-6.125&& 298 & 15.650&-7.952&-8.239&-8.125\\
        &288 & 13.705&-5.479&-5.772&-5.610 && 300 & 15.313&-7.440&-7.711&-7.607\\
        & 290 & 13.676&-5.490&-5.753&-5.662 && 302 & 14.782&-6.558&-6.821&-6.685\\
          &292 &  13.441&-5.086&-5.328&-5.387&& 304 & 14.914&-6.865&-7.092&-6.905\\
         & 294 &  13.215&-4.686&-4.907&-4.907&& 306 & 14.667&-6.479&-6.685&-6.453\\
         & 296 &  13.316&-4.964& -5.148 &-5.245&& 308 & 14.644&-6.504&-6.678&-6.453\\
        & 298 & 12.981&-4.324&-4.492 &-4.493&& 310 & 15.412&-7.945&-8.054&-7.774\\
         & 300 & 13.294&-5.055&-5.174&-5.176&& 312 & 13.833&-5.066&-5.213&-4.998\\
          &302& 12.866&-4.211&-4.318&-4.346&& 314& 13.223&-3.864&-4.007&-4.185\\
         & 304 & 12.740&-4.003&-4.082&-4.020&& 316 & 13.175&-3.827&-3.940&-4.147\\
         & 306 & 13.765&-6.216& -6.210&-6.156 && 318 & 12.544&-2.469&-2.583&-2.824\\
         & 308 &  12.945&-4.591&-4.591&-4.556 && 320 & 11.889&-0.939&-1.057&-1.350\\
         & 310 & 11.478&-1.189&-1.236&-1.297 && 322& 12.233&-1.873&-1.935&-2.509\\
          &312 &11.196&-0.518&-0.548&-0.921&& 324 & 11.975&-1.303&-1.346&-1.949\\
         & 314 &  10.739&0.679&0.655 &0.233&& 326 & 12.241&-2.029&-2.023&-2.652\\
         & 316 & 9.173&5.631& 5.526&4.360&& 328 & 11.117&0.826&0.799&-0.199\\
      & 318 &  9.912&3.045&3.034 &2.243&& 330& 12.311&-2.335&-2.256&-2.941\\
        & 320 &  9.660&3.806&3.810&2.978 & &  & &&&\\
    \end{tabular}
    }
    \label{tab9}
    \end{ruledtabular}
\end{table*}

\bibliography{study}

@article{qu2014comparative,
  title={Comparative studies of Coulomb barrier heights for nuclear models applied to sub-barrier fusion},
  author={Qu, WW and Zhang, GL and Zhang, HQ and Wolski, R},
  journal={Physical Review C},
  volume={90},
  number={6},
  pages={064603},
  year={2014},
  publisher={APS}
}

@article{zdeb2013half,
  title={Half-lives for $\alpha$ and cluster radioactivity within a Gamow-like model},
  author={Zdeb, A and Warda, M and Pomorski, K},
  journal={Physical Review C—Nuclear Physics},
  volume={87},
  number={2},
  pages={024308},
  year={2013},
  publisher={APS}
}

@article{sun2017systematic,
  title={Systematic study of $\alpha$ decay half-lives of doubly odd nuclei within the two-potential approach},
  author={Sun, Xiao-Dong and Deng, Jun-Gang and Xiang, Dong and Guo, Ping and Li, Xiao-Hua},
  journal={Physical Review C},
  volume={95},
  number={4},
  pages={044303},
  year={2017},
  publisher={APS}
}

@article{santhosh2018alpha,
  title={$\alpha$-decay half-lives of superheavy nuclei from a modified generalized liquid-drop model},
  author={Santhosh, KP and Nithya, C and Hassanabadi, H and Akrawy, Dashty T},
  journal={Physical Review C},
  volume={98},
  number={2},
  pages={024625},
  year={2018},
  publisher={APS}
}

@article{oganessian2010synthesis,
  title={Synthesis of a new element with atomic number Z= 117},
  author={Oganessian, Yu Ts and Abdullin, F Sh and Bailey, PD and Benker, DE and Bennett, ME and Dmitriev, SN and Ezold, Julie G and Hamilton, JH and Henderson, Roger A and Itkis, MG and others},
  journal={Physical review letters},
  volume={104},
  number={14},
  pages={142502},
  year={2010},
  publisher={APS}
}

@article{oganessian2007heaviest,
  title={Heaviest nuclei from 48Ca-induced reactions},
  author={Oganessian, Yuri},
  journal={Journal of Physics G: Nuclear and Particle Physics},
  volume={34},
  number={4},
  pages={R165},
  year={2007},
  publisher={IOP Publishing}
}

@article{hosseini2017theoretical,
  title={Theoretical approaches to alpha decay half-lives of super-heavy nuclei},
  author={Hosseini, SS and Hassanabadi, H},
  journal={Chinese Physics C},
  volume={41},
  number={6},
  pages={064101},
  year={2017},
  publisher={IOP Publishing}
}

@article{javadimanesh2013investigation,
  title={Investigation of deformed nuclei with a new potential combination},
  author={Javadimanesh, E and Hassanabadi, H and Rajabi, AA and Rahimov, H and Zarrinkamar, S},
  journal={Chinese Physics C},
  volume={37},
  number={11},
  pages={114102},
  year={2013},
  publisher={IOP Publishing}
}

@article{hosseini2019alpha,
  title={Alpha particle preformation factor of spherical nuclei for $67 \leq z \leq 91$},
  author={Hosseini, SS and Hassanabadi, H and Akrawy, Dashty T},
  journal={Modern Physics Letters A},
  volume={34},
  number={05},
  pages={1950039},
  year={2019},
  publisher={World Scientific}
}

@article{hodgson2003cluster,
  title={Cluster emission, transfer and capture in nuclear reactions},
  author={Hodgson, Peter Edward and B{\v{e}}t{\'a}k, E},
  journal={Physics reports},
  volume={374},
  number={1},
  pages={1--89},
  year={2003},
  publisher={Elsevier}
}

@article{geiger1911lvii,
  title={LVII. The ranges of the $\alpha$ particles from various radioactive substances and a relation between range and period of transformation},
  author={Geiger, Hans and Nuttall, JM},
  journal={The London, Edinburgh, and Dublin Philosophical Magazine and Journal of Science},
  volume={22},
  number={130},
  pages={613--621},
  year={1911},
  publisher={Taylor \& Francis}
}

@article{royer2000alpha,
  title={Alpha emission and spontaneous fission through quasi-molecular shapes},
  author={Royer, Guy},
  journal={Journal of Physics G: Nuclear and Particle Physics},
  volume={26},
  number={8},
  pages={1149},
  year={2000},
  publisher={IOP Publishing}
}

@article{akrawy2018new,
  title={New empirical formula for $\alpha$-decay calculations},
  author={Akrawy, Dashty T and Ahmed, Ali H},
  journal={International Journal of Modern Physics E},
  volume={27},
  number={08},
  pages={1850068},
  year={2018},
  publisher={World Scientific}
}

@article{akrawy2019alpha,
  title={$\alpha$-decay systematics for superheavy nuclei},
  author={Akrawy, Dashty T and Ahmed, Ali H},
  journal={Physical Review C},
  volume={100},
  number={4},
  pages={044618},
  year={2019},
  publisher={APS}
}

@article{qi2009universal,
  title={Universal decay law in charged-particle emission and exotic cluster radioactivity},
  author={Qi, Chong and Xu, FR and Liotta, Roberto J and Wyss, Ramon},
  journal={Physical review letters},
  volume={103},
  number={7},
  pages={072501},
  year={2009},
  publisher={APS}
}

@article{viola1966nuclear,
  title={Nuclear systematics of the heavy elements—II Lifetimes for alpha, beta and spontaneous fission decay},
  author={Viola Jr, VE and Seaborg, GT},
  journal={Journal of Inorganic and Nuclear Chemistry},
  volume={28},
  number={3},
  pages={741--761},
  year={1966},
  publisher={Elsevier}
}

@article{denisov2024empirical,
  title={Empirical relations for $\alpha$-decay half-lives: The effect of deformation of daughter nuclei},
  author={Denisov, V Yu},
  journal={Physical Review C},
  volume={110},
  number={1},
  pages={014604},
  year={2024},
  publisher={APS}
}

@article{guo2015nuclear,
  title={The nuclear deformation and the preformation factor in the $\alpha$-decay of heavy and superheavy nuclei},
  author={Guo, Shuqing and Bao, Xiaojun and Gao, Yuan and Li, Junqing and Zhang, Hongfei},
  journal={Nuclear Physics A},
  volume={934},
  pages={110--120},
  year={2015},
  publisher={Elsevier}
}

@article{zhang2006alpha,
  title={$\alpha$ decay half-lives of new superheavy nuclei within a generalized liquid drop model},
  author={Zhang, Hongfei and Zuo, Wei and Li, Junqing and Royer, Guy},
  journal={Physical Review C—Nuclear Physics},
  volume={74},
  number={1},
  pages={017304},
  year={2006},
  publisher={APS}
}

@article{zanganah2020calculation,
  title={Calculation of $\alpha$-decay and cluster half-lives for 197--226fr using temperature-dependent proximity potential model},
  author={Zanganah, V and Akrawy, Dashty T and Hassanabadi, H and Hosseini, SS and Thakur, Shagun},
  journal={Nuclear Physics A},
  volume={997},
  pages={121714},
  year={2020},
  publisher={Elsevier}
}

@article{yahya2020alpha,
  title={Alpha decay half-lives of 171-189Hg isotopes using Modified Gamow-like model and temperature dependent proximity potential},
  author={Yahya, WA},
  journal={Journal of the Nigerian Society of Physical Sciences},
  pages={250--256},
  year={2020}
}

@article{gurvitz1987decay,
  title={Decay width and the shift of a quasistationary state},
  author={Gurvitz, SA and Kalbermann, G},
  journal={Physical review letters},
  volume={59},
  number={3},
  pages={262},
  year={1987},
  publisher={APS}
}

@article{moghaddari2020influence,
  title={Influence of the Pauli exclusion principle on $\alpha$ decay},
  author={Moghaddari Amiri, M and Ghodsi, ON},
  journal={Physical Review C},
  volume={102},
  number={5},
  pages={054602},
  year={2020},
  publisher={APS}
}

@article{jian2009alpha,
  title={$\alpha$-decay half-lives of superheavy nuclei and general predictions},
  author={Jian-Min, Dong and Hong-Fei, Zhang and Yan-Zhao, Wang and Wei, Zuo and Xin-Ning, Su and Jun-Qing, Li},
  journal={Chinese Physics C},
  volume={33},
  number={8},
  pages={633},
  year={2009},
  publisher={IOP Publishing}
}

@article{yan2009branching,
  title={Branching ratios of $\alpha$ decay for nuclei near deformed shell closures},
  author={Yan-Zhao, Wang and Hong-Fei, Zhang and Jian-Min, Dong and Xin-Ning, Su and Wei, Zuo and Jun-Qing, Li},
  journal={Chinese Physics Letters},
  volume={26},
  number={6},
  pages={062101},
  year={2009},
  publisher={IOP Publishing}
}

@article{ren2012new,
  title={New Geiger-Nuttall law for $\alpha$ decay of heavy nuclei},
  author={Ren, Yuejiao and Ren, Zhongzhou},
  journal={Physical Review C—Nuclear Physics},
  volume={85},
  number={4},
  pages={044608},
  year={2012},
  publisher={APS}
}

@article{koura2012phenomenological,
  title={Phenomenological formula for alpha-decay half-lives},
  author={Koura, Hiroyuki},
  journal={Journal of nuclear science and technology},
  volume={49},
  number={8},
  pages={816--823},
  year={2012},
  publisher={Taylor \& Francis}
}

@article{deng2020improved,
  title={Improved empirical formula for $\alpha$-decay half-lives},
  author={Deng, Jun-Gang and Zhang, Hong-Fei and Royer, Guy},
  journal={Physical Review C},
  volume={101},
  number={3},
  pages={034307},
  year={2020},
  publisher={APS}
}

@article{saxena2021new,
  title={A new empirical formula for $\alpha$-decay half-life and decay chains of Z= 120 isotopes},
  author={Saxena, G and Jain, A and Sharma, PK},
  journal={Physica Scripta},
  volume={96},
  number={12},
  pages={125304},
  year={2021},
  publisher={IOP Publishing}
}

@article{soylu2021extended,
  title={Extended universal decay law formula for the $\alpha$ and cluster decays},
  author={Soylu, As{\i}m and Qi, Chong},
  journal={Nuclear Physics A},
  volume={1013},
  pages={122221},
  year={2021},
  publisher={Elsevier}
}

@article{akrawy2022alpha,
  title={$\alpha$-decay half-lives new semi-empirical relationship including asymmetry, angular momentum and shell effects},
  author={Akrawy, Dashty T and Poenaru, Dorin N and Ahmed, Ali H and Sihver, Lembit},
  journal={Nuclear Physics A},
  volume={1021},
  pages={122419},
  year={2022},
  publisher={Elsevier}
}

@article{akrawy2017alpha,
  title={Alpha decay calculations with a new formula},
  author={Akrawy, Dashty T and Poenaru, DN},
  journal={Journal of Physics G: Nuclear and Particle Physics},
  volume={44},
  number={10},
  pages={105105},
  year={2017},
  publisher={IOP Publishing}
}

@article{akrawy2022generalization,
  title={Generalization of the screened universal $\alpha$-decay law by asymmetry and angular momentum},
  author={Akrawy, Dashty T and Budaca, AI and Saxena, G and Ahmed, Ali H},
  journal={The European Physical Journal A},
  volume={58},
  number={8},
  pages={145},
  year={2022},
  publisher={Springer}
}

@article{yu2009model,
  title={Model investigation on the probability of QGP formation at different centralities in relativistic heavy ion collisions},
  author={Yu, Meiling and Xu, Mingmei and Liu, Zhengyou and Liu, Lianshou},
  journal={Physical Review C—Nuclear Physics},
  volume={80},
  number={6},
  pages={064908},
  year={2009},
  publisher={APS}
}

@article{denisov2010erratum,
  title={Erratum: $\alpha$ decay of even-even superheavy elements [Phys. Rev. C 81, 034613 (2010)]},
  author={Denisov, V Yu and Khudenko, AA},
  journal={Physical Review C—Nuclear Physics},
  volume={82},
  number={5},
  pages={059903},
  year={2010},
  publisher={APS}
}

@article{deng2017improved,
  title={Improved double-folding $\alpha$-nucleus potential by including nuclear medium effects},
  author={Deng, Daming and Ren, Zhongzhou},
  journal={Physical Review C},
  volume={96},
  number={6},
  pages={064306},
  year={2017},
  publisher={APS}
}

@article{ismail2017alpha,
  title={Alpha-decay of deformed superheavy nuclei as a probe of shell closures},
  author={Ismail, M and Seif, WM and Adel, A and Abdurrahman, A},
  journal={Nuclear Physics A},
  volume={958},
  pages={202--210},
  year={2017},
  publisher={Elsevier}
}

@article{wang2014surface,
  title={Surface diffuseness correction in global mass formula},
  author={Wang, Ning and Liu, Min and Wu, Xizhen and Meng, Jie},
  journal={Physics Letters B},
  volume={734},
  pages={215--219},
  year={2014},
  publisher={Elsevier}
}

@article{xu2022unified,
  title={A unified formula for $\alpha$ decay half-lives},
  author={Xu, Yang-Yang and Zhu, De-Xing and Chen, Xun and Wu, Xi-Jun and He, Biao and Li, Xiao-Hua},
  journal={The European Physical Journal A},
  volume={58},
  number={9},
  pages={163},
  year={2022},
  publisher={Springer}
}

@article{luo2023improved,
  title={An improved empirical formula of $\alpha$ decay half-lives for superheavy nuclei},
  author={Luo, Song and Qi, Lin-Jing and Zhang, Dong-Meng and He, Biao and Chu, Peng-Cheng and Li, Xiao-Hua},
  journal={The European Physical Journal A},
  volume={59},
  number={6},
  pages={125},
  year={2023},
  publisher={Springer}
}

@article{manjunatha2018investigations,
  title={Investigations of the synthesis of the superheavy element z= 122},
  author={Manjunatha, HC and Sridhar, KN and Sowmya, N},
  journal={Physical Review C},
  volume={98},
  number={2},
  pages={024308},
  year={2018},
  publisher={APS}
}

@article{anushree2025entrance,
  title={Entrance channel-dependent compound nucleus formation probability of heavy nuclei},
  author={Anushree, HS and Shubha, S and Manjunatha, HC and Sowmya, N},
  journal={Pramana},
  volume={99},
  number={2},
  pages={1--8},
  year={2025},
  publisher={Springer}
}

@article{madhu2024cr,
  title={Cr-induced fusion reactions to synthesize superheavy elements},
  author={Madhu, S and Manjunatha, HC and Sowmya, N and Rajesh, BM and Seenappa, L and Susheela, RS},
  journal={Nuclear Science and Techniques},
  volume={35},
  number={5},
  pages={90},
  year={2024},
  publisher={Springer}
}

@article{manjunatha2024effect,
  title={Effect of Quadrupole and Hexadecapole Deformations of Target on Projectile},
  author={Manjunatha, Narayanaswamy and Manjunatha, Holaly Chandrashekara Shastry and Sowmya, Nagaraj and Sridhar, Krishnachari Nagarathnamma and Prabhavathi, Padhmaghatta Somashekaraih},
  journal={Journal of the Physical Society of Japan},
  volume={93},
  number={5},
  pages={054201},
  year={2024},
  publisher={The Physical Society of Japan}
}

@article{sowmya2024optimal,
  title={Optimal incident energy of heavy ion fusion},
  author={Sowmya, N and Manjunatha, HC and Sridhar, KN and Armstrong Arasu, MM},
  journal={Physical Review C},
  volume={109},
  number={2},
  pages={024610},
  year={2024},
  publisher={APS}
}

@article{kondev2021nubase2020,
  title={The NUBASE2020 evaluation of nuclear physics properties},
  author={Kondev, FG and Wang, Meng and Huang, WJ and Naimi, S and Audi, G},
  journal={Chinese Physics C},
  volume={45},
  number={3},
  pages={030001},
  year={2021},
  publisher={IOP Publishing}
}

@article{ismail2009effect,
  title={Effect of octupole and higher deformations on Coulomb barrier},
  author={Ismail, M and Seif, WM and Botros, MM},
  journal={Nuclear Physics A},
  volume={828},
  number={3-4},
  pages={333--347},
  year={2009},
  publisher={Elsevier}
}

@article{xiao2023half,
  title={Half-lives for proton emission and $\alpha$ decay within the deformed Gamow-like model},
  author={Xiao, Qiong and Cheng, Jun-Hao and Wang, Bing-Lin and Xu, Yang-Yang and Zou, You-Tian and Yu, Tong-Pu},
  journal={Journal of Physics G: Nuclear and Particle Physics},
  volume={50},
  number={8},
  pages={085102},
  year={2023},
  publisher={IOP Publishing}
}
\end{document}